\documentclass[12pt]{spieman}  
\usepackage{amsmath,amsfonts,amssymb}
\usepackage{graphicx}
\usepackage{setspace}
\usepackage{subcaption}
\usepackage{tocloft}
\usepackage{xcolor}
\usepackage{soul}

\title{End-to-end Deep Learning Pipeline for Microwave Kinetic Inductance Detector (MKID) Resonator Identification and Tuning}

\author[a,*]{Neelay Fruitwala}
\author[b]{Alex B Walter}
\author[a]{John I Bailey, III }
\author[a]{Rupert Dodkins}
\author[a]{Benjamin A Mazin}

\affil[a]{Department of Physics, University of California, Santa Barbara, California 93106, USA}
\affil[b]{Jet Propulsion Laboratory; California Institute of Technology, Pasadena, California 91125, USA}

\cftpagenumbersoff{figure}
\cftpagenumbersoff{table} 
\begin{document} 
\maketitle

\begin{abstract}
We present the development of a machine learning based pipeline to fully automate the calibration of the frequency comb used to read out optical/IR Microwave Kinetic Inductance Detector (MKID) arrays. This process involves determining the resonant frequency and optimal drive power of every pixel (i.e. resonator) in the array, which is typically done manually. Modern optical/IR MKID arrays, such as DARKNESS (DARK-speckle Near-infrared Energy-resolving Superconducting Spectrophotometer) and MEC (MKID Exoplanet Camera), contain 10-20,000 pixels, making the calibration process extremely time consuming; each 2000 pixel feedline requires 4-6 hours of manual tuning. Here we present a pipeline which uses a single convolutional neural network (CNN) to perform both resonator identification and tuning simultaneously. We find that our pipeline has performance equal to that of the manual tuning process, and requires just twelve minutes of computational time per feedline.
\end{abstract}


{\noindent \footnotesize\textbf{*}Neelay Fruitwala,  \linkable{neelay@ucsb.edu} }

\begin{spacing}{2}   

\section{Introduction}
\label{sect:intro}  
Microwave kinetic inductance detectors (MKIDs) are a cryogenically cooled, superconducting detector technology with applications in astronomical observations in the submilimeter through visible wavelengths. In the optical and IR, MKIDs have zero read noise, single photon sensitivity, microsecond time resolution, and are energy resolving ($R \approx 6$) \cite{paul2017}. The current generation of optical/IR MKIDs is focused on high contrast imaging, where high sensitivity/low noise are required for observing faint companions, and energy and time resolution are useful for resolving atmospheric effects (which are the dominant noise source for such observations) \cite{}. Two MKID IFUs have been built for this purpose: MEC, a 20,000 pixel camera behind SCExAO at Subaru Observatory \cite{mec}, and DARKNESS, a 10,000 pixel camera behind P3K (PALM-3000 AO system) and SDC (stellar double coronagraph) at Palomar Observatory \cite{darkness}.

In an MKID detector array, each pixel is a superconducting LC resonant circuit with a photosensitive inductor; when a photon hits the inductor, its inductance changes, changing the resonant frequency of the circuit. This change can be measured by driving the resonator with an on-resonance probe tone and continuosly monitoring its phase response. MKIDs are highly multiplexible; each pixel can be tuned to a different resonant frequency, allowing large numbers of resonators to be placed in parallel on a single microwave feedline. The resonators on this feedline are then read out using a comb of probe tones, each one tuned to the frequency of a single resonator. The current generation of MKID instruments have 2000 resonators per feedline within a 4-8 GHz bandwidth \cite{paul2017}. The calibration of this probe tone frequency comb is the subject of this paper. 

This calibration process generally involves two stages: 1) identification of resonator frequencies; and 2) tuning the drive power of each resonator probe tone. Due to imperfections in the array fabrication process, resonator frequencies deviate significantly more (from their design values) than the average resonator frequency spacing, and must be measured empirically. Additionally, the photon detection SNR of each resonator is very sensitive to the power of its input probe tone \cite{swenson, zobrist}. This ``ideal" drive power varies considerably across the array and must be tuned individually for each resonator.

The calibration process requires significant manual intervention and can be prohibitively time consuming; calibrating a single 2000 pixel feedline can take 4-6 hours. Because calibration is required to fully characterize an MKID array, this process can present a significant bottleneck for testing new devices.

Machine learning has been shown to be a promising method for tuning resonator drive power \cite{dodkins}. However, using machine learning to tune power alone (and not frequency) suffers from a few pitfalls, and we find that manual tuning is still required to refine the ML selected powers. Building on this earlier work, we present a unified deep learning framework for performing both frequency and power calibration steps simultaneously. We find that our method has performance equal to that of manual tuning, and reduces calibration time to less than 15 minutes per feedline.

\section{Resonator Identification and Tuning: Baseline Methodology}
\label{sec:baseline_method}

The standard procedure for MKID array calibration treats frequency identification and drive power tuning independently. We outline this procedure below, then explain some of the issues that require manual tuning to address.

\begin{figure}[h]
    \centering
    \includegraphics[scale=0.65, trim={60 0 0 40}, clip]{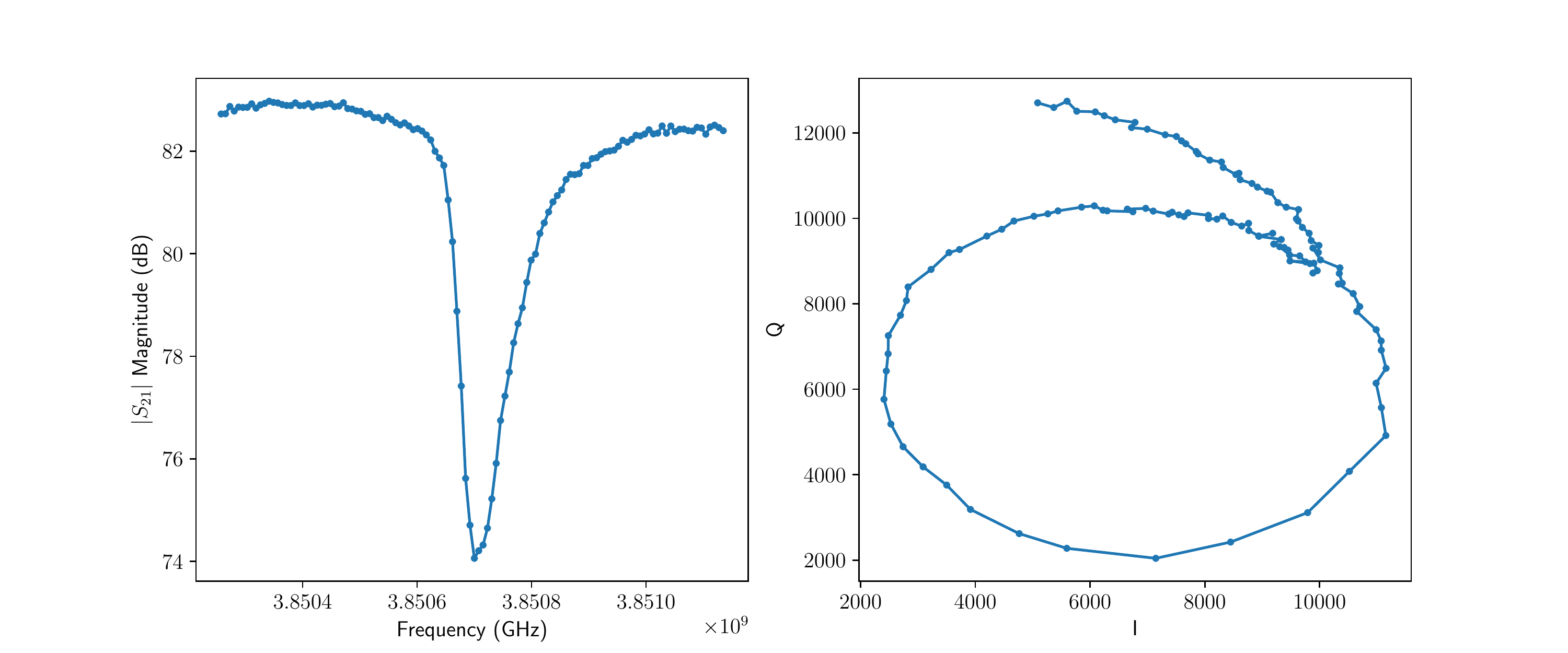}
    \caption{Resonators absorb power in a $\approx 200$ kHz band around the resonant frequency (hence a dip in $|S_{21}|$; left), but leave tones outside this band relatively unmodified. If $S_{21}(f)$ is plotted in the complex plane over this frequency band, the resonator will trace out a ``loop" (right).}
    \label{fig:goodloop}
\end{figure}

All calibration steps are performed using measurements of the complex frequency response $S_{21}(f) = I(f) + iQ(f)$ of the MKID array in the resonator frequency band (approximately 4-8 GHz for the current generation of instruments). When $S_{21}(f)$ is plotted on the complex plane, resonators manifest as ``loops" when driven at the correct power (figure \ref{fig:goodloop}). Resonators absorb power close to resonance, so they also correspond to local minima in $|S_{21}|$. Frequency response data is taken over a wide range of drive powers for the tuning stage.

\subsection{Resonator Identification}
\label{sec:baseline_res_id}

Due to imperfections in the MKID array fabrication process, there is an ${\approx}50$ MHz spread between the nominal resonator design frequency and the actual measured frequency. Since pixels are designed to be separated by ${\approx}2$ MHz, this spread makes the nominal design frequency essentially useless for identifying resonators. Instead, we measure the magnitude of the frequency response ($|S_{21}|$) across the full 4-8 GHz bandwidth of each microwave feedline (``widesweep''), where resonators appear as narrow 5-20 dB deep "dips" in transmission (figure \ref{fig:goodloop}). These dips can be accurately flagged automatically using a spatial bandpass filter (to select for the band corresponding to resonator width in the $|S_{21}(f)|$ signal) followed by a peak-detection algorithm such as SciPy's signal.find\_peaks function \cite{scipy}. 

\subsection{Resonator Drive Power Tuning}
\label{sec:baseline_power}

\begin{figure}[h]
    \centering
    \includegraphics[scale=.75, trim={35 0 0 40}, clip]{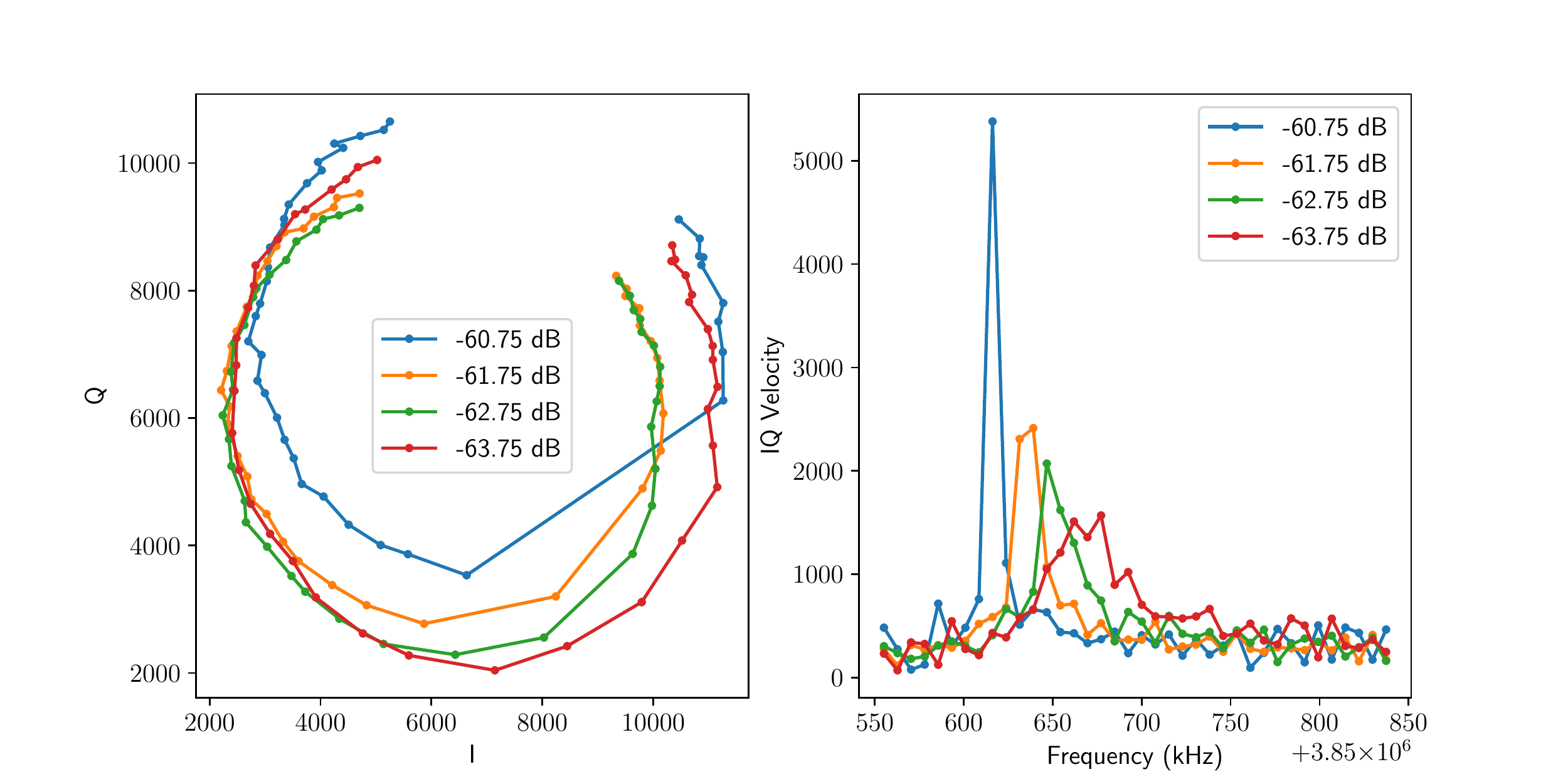}
    \caption{Complex frequency response $S_{21}(f)$ (left) and corresponding IQV (right) of a single resonator measured at four different powers. The resonator is clearly bifurcated when driven at -60.75 dB, as indicated by the discontinuity in the resonator loop. There is still some apparent discontinuity at -61.75 dB; -62.75 dB or -63.75 dB would be reasonable drive powers for this resonator. Power is measured relative to the DAC full scale output.}
    \label{fig:mult_power_loops}
\end{figure}

In order to maximize photon detection SNR, resonators should be driven at the highest possible power, as amplifier induced phase noise decreases linearly with drive power \cite{zobrist, readout}. However, if a resonator is driven too hard, it can enter a bistable ``bifurcation'' state, rendering it useless for photon detection \cite{swenson}. So, we take the ``ideal'' resonator drive power to be slightly (1-3 dB) below the bifurcation state. For most resonators, the bifurcation state can easily be identified in the complex frequency response of the resonator across its linewidth ($\approx 200\  \textrm{kHz}$) (see figure \ref{fig:mult_power_loops}), making it relatively straightforward for a human operator (or machine learning algorithm) to identify the correct drive power.

To tune each resonator, we select a 450-750 kHz window around the frequency dip identified in the previous step, and measure the complex (I and Q) frequency response in this window at a variety of drive powers (typically a 31 dB range in steps of 1 dB). We then feed this data into a neural net classifier that estimates the drive power \cite{dodkins}.

To refine the estimate of optimal resonator drive frequency at a given power, we define the quantity IQ-velocity (IQV), which is the magnitude of the derivative of $S_{21}$ in the complex plane wrt frequency:
\begin{equation}
    IQV(f_i) = |\Delta_f S_{21}(f_i)| = \sqrt{[I(f_{i+1}) - I(f_i)]^2 + [Q(f_{i+1}) - Q(f_i)]^2}
\end{equation}
where $i$ indexes frequency, and all $S_{21}$ measurements are performed at a single power. The point along a resonator loop with maximum IQV is the ideal drive frequency, since it is where the phase response of the resonator is most sensitive to changes in frequency. This is often different from the $|S_{21}(f)|$ minimum (figure \ref{issue2plots}). So, after the optimal drive power is estimated, the frequency identified in the previous stage (section \ref{sec:baseline_res_id}) is changed to the IQV maximum at the estimated power. This is typically done using a peak finding algorithm in conjuction with a boxcar smoothing filter.

After automated power and frequency determination, a human operator may look through the data set to refine the neural network's initial guesses. 

If two resonators are separated by less than $200$ kHz, we only read out one of them to prevent their probe tones from interfering \cite{readout}. We choose the higher frequency resonator to prevent undesirable resonator crosstalk during photon detection events \footnote{Since the resonant frequency decreases during detection events,  the higher frequency resonator might enter the readout band of the lower frequency resonator, interfering with that resonator's probe tone.}. After the identification and tuning process, the final resonator list is pruned according to this criteria; we define the resulting list to be the set of ``usable resonators''.

\subsection{Problems with the Baseline Methodology}

We find that the automated steps of the above method are inadequate for providing an accurate calibration solution; 15-20\% of resonators may be miscalibrated. Most of these calibration errors are related to choosing a frequency window around each identified resonator (section \ref{sec:baseline_res_id}) to feed into the neural network power classifier. We detail these issues below, using an analysis of a single representative feedline from the MEC array ``Hypatia'' (call this feedline ``Hypatia6'').

\begin{figure}[h]
    \includegraphics[scale=0.99, trim={0 0 0 0}, clip]{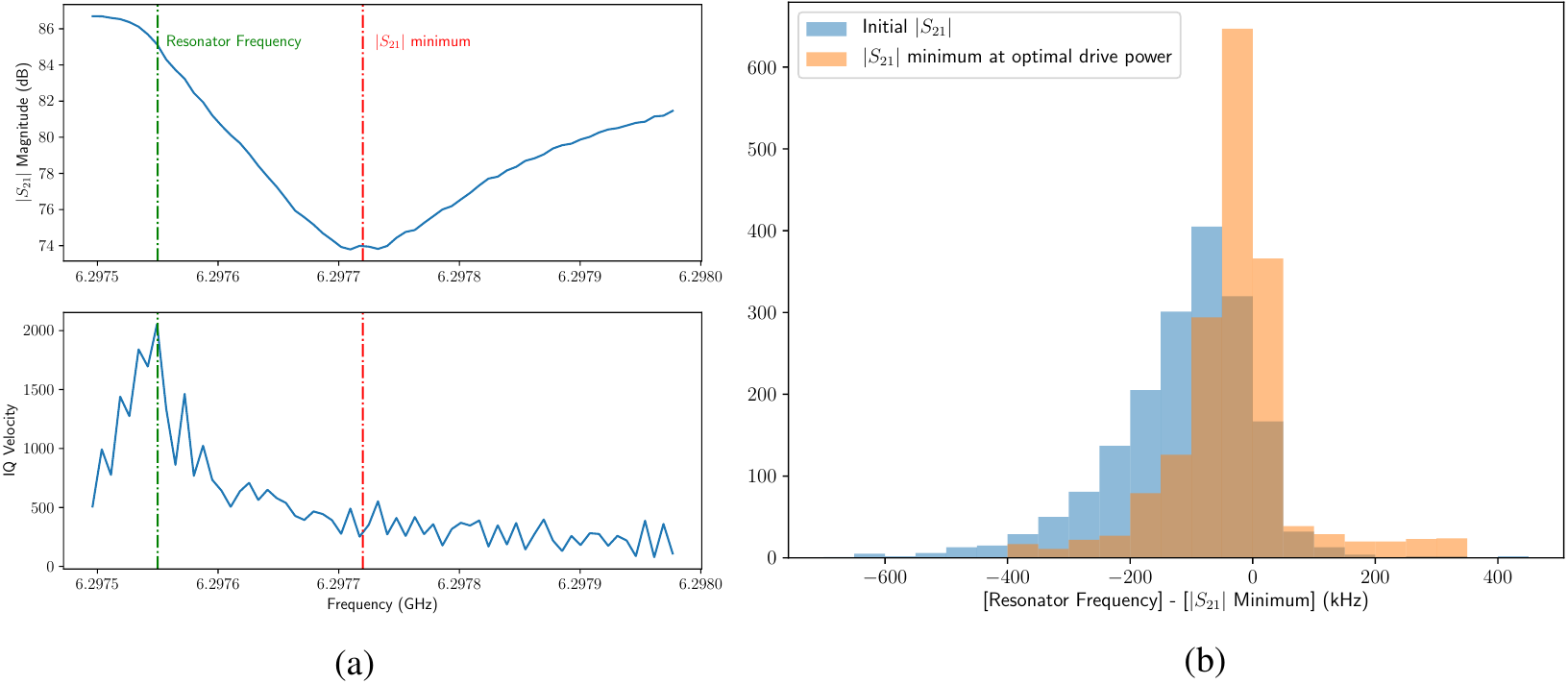}
    \caption{Illustration of issue (2). a) Plot of $|S_{21}|$ and IQ velocity for a single resonator, measured at its optimal drive power. For this resonator, the $|S_{21}|$ minimum and IQ velocity maximum (taken to be the ``actual" resonator frequency) are separated by 170 kHz. b) Histograms of this discrepancy across Hypatia6. Initial $|S_{21}|$ (blue) shows the difference between the actual resonator frequency and that flagged in the identification step. Some of this is due to the resonator frequency changing with power, and can be corrected for to some extent by re-centering the machine learning window accordingly (orange).}
    \label{issue2plots}

\end{figure}

Choosing a frequency window around each resonator is complicated by the following factors: 1) Resonators often move significantly in frequency space as a function of power (figure \ref{fig:mult_power_loops}); and 2) The local minimum in $|S_{21}|$ that is used to identify the resonator is often not aligned with the IQV maximum; this discrepancy can be larger than the linewidth of the resonator (figure \ref{issue2plots}). 

Issue (1) can be dealt with to some extent by re-centering the window around the $|S_{21}|$ dip at each power. However, this is complicated by issue (2), since the $|S_{21}|$ dip doesn't always shift by the same amount as the actual resonator frequency.

\begin{figure}[h]
    \includegraphics[]{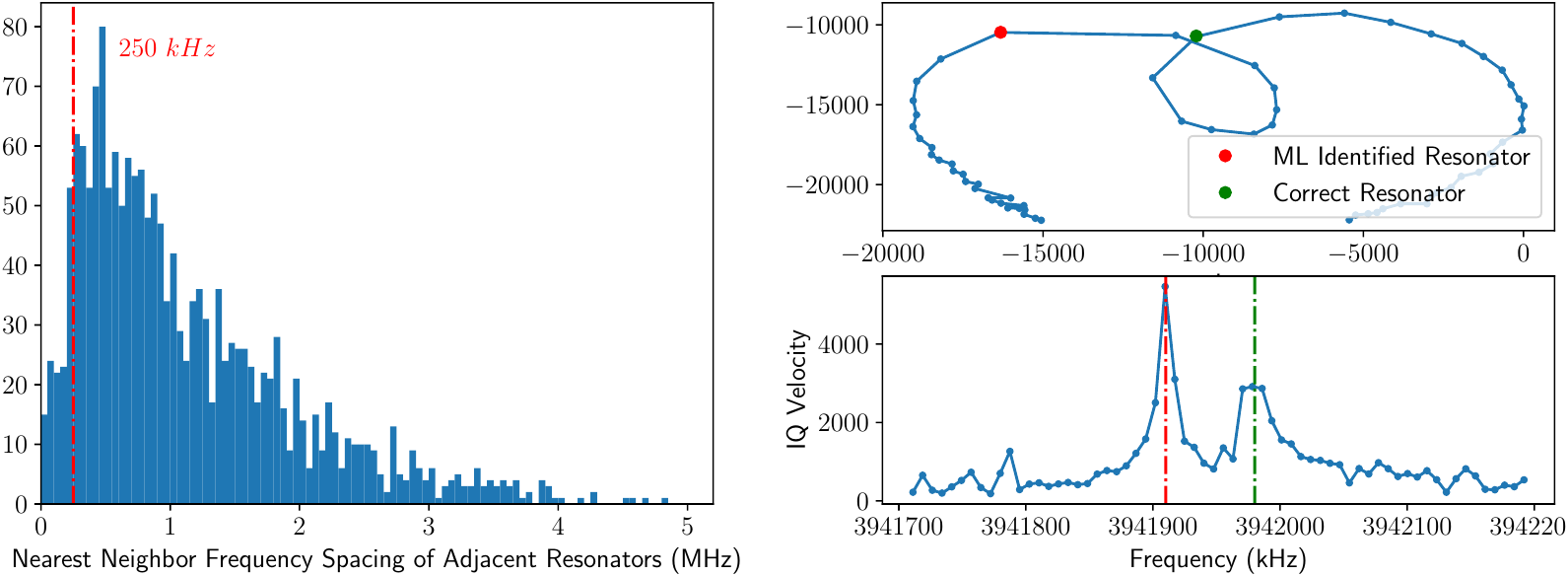}
    \caption{left: histogram of the frequency spacing between resonator and it's nearest neighbor in frequency space for Hypatia6 \protect\footnotemark. Approximately 8\% of resonators contain another resonator within a 500 kHz window; right: example of a window with multiple resonators. In this case, the machine learning algorithm selected the correct power for the higher frequency resonator (which was also the one flagged during the identification stage), but selected the frequency of the lower frequency resonator (which is bifurcated at this power).}
    \label{doublesplots}

\end{figure}

Because of issue (2), we find that windows must be quite wide; e.g. for the Hypatia6 feedline, a 500 kHz window will only include ${\approx}90\%$ of the resonators \footnote{The median resonator linewidth is 140 kHz, so to be included in a 500 kHz window, we assume that the resonator frequency is $\pm 180\ \textrm{kHz}$ from the window center.}. Using larger windows however increases the chance that there will be multiple resonators within the same window; which degrades the capability of the machine learning algorithm to tune the resonator (figure \ref{doublesplots}).

We find that while the neural network power classifier generally has good performance on standalone resonators, manual intervention is required when there are multiple resonators within the same window, or the resonator is too far from the frequency identified in the previous stage and lies outside the window. For the Hypatia6 feedline, this comprises 15-20\% of resonators. It is difficult to determine a priori which resonators meet these criteria (particularly when the resonator lies outside the initial window), so manual inspection is required for all resonators in order to fully optimize array performance. This is a time consuming process, requiring 4-6 hours per feedline. To address these issues, we present an approach for doing both frequency identification and tuning in a single step using the same convolutional neural network. 

\footnotetext{This histogram includes only resonators satisfying the ``usable resonator'' criterion defined in section \protect\ref{sec:baseline_power}. However, we allow resonators that violate this criterion to count as nearest neighbors, since they may still interfere with the tuning process even if they aren't in the final resonator list.}

\section{End-to-end Machine Learning: System Overview}

\subsection{Overall Architecture}

The full $S_{21}$ dataset can be conceptualized as a 2D ``image", with a real ($I$) and imaginary ($Q$) value at each point in $power \times freq$ space. Each resonator at its optimal drive power and resonant frequency is located at a single point in this image, and its influence extends to a small region surrounding this point. Resonator identification/tuning can then be posed as an object detection problem in $power \times freq$ space. 

One of the simplest methods of implementing machine learning based object detection is to use a sliding window classifier. In this scheme, we take a small window around each point in the $S_{21}(p, f)$ image and feed it into a deep convolutional neural network (CNN) \cite{deeplearning}, which assigns the window a set of scores representing the likelihood of it being a member of each of the following classes:

\begin{enumerate}
        \item window is centered around a resonator at the correct power and frequency
        \item window is centered around a bifurcated (saturated) resonator
        \item window is centered around an underpowered resonator
        \item there is no resonator inside the window
\end{enumerate}

The output of the classification stage is four $N_{power} \times N_{freq}$ images, each containing the classification score at each point for one of the above classes. To find resonators, we use a relatively simple criteria to select $N$ local maxima above some threshold of the image representing the scores for class (1) (detailed in section \ref{sec:nn_resid}).

\begin{figure}[h]
    \centering
    \includegraphics[scale=0.7, trim={20 500 10 50},  clip]{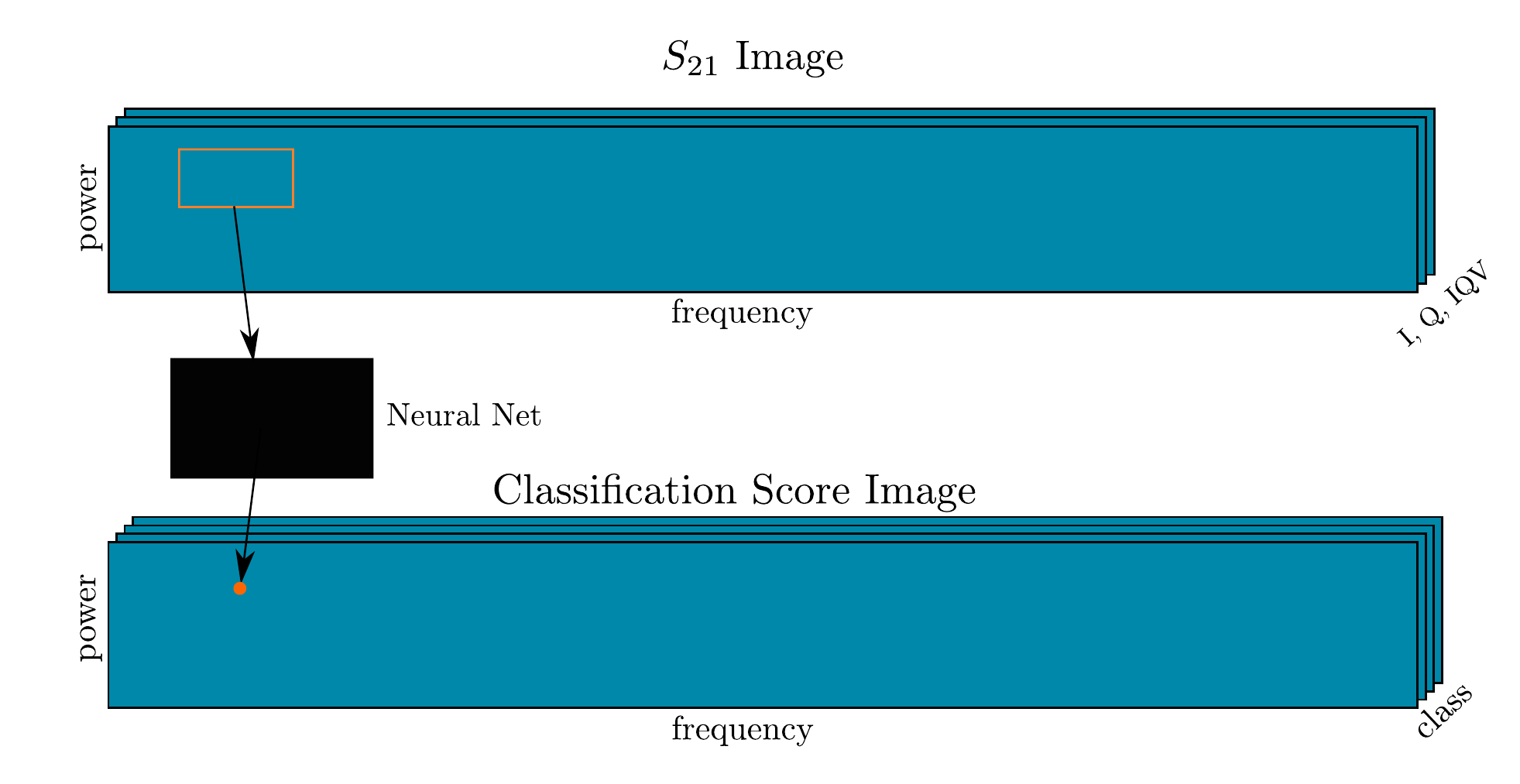}
    \caption{Schematic of overall architecture. The input data set is a 3-color image (one for each of $I$, $Q$, and $IQV$) containing the frequency response of the MKID feedline at multiple powers. A small window is generated around each $(p, f)$ point, and fed into a convolutional neural network to generate a set of classification scores for that point. This results in a $N_{power} \times N_{freq}$ output image with four colors, one for each classification score.}
    \label{fig:block_diagram}
\end{figure}

\begin{figure}[h]
    \centering
    \includegraphics[scale=0.45, trim={22 0 18 27}, clip]{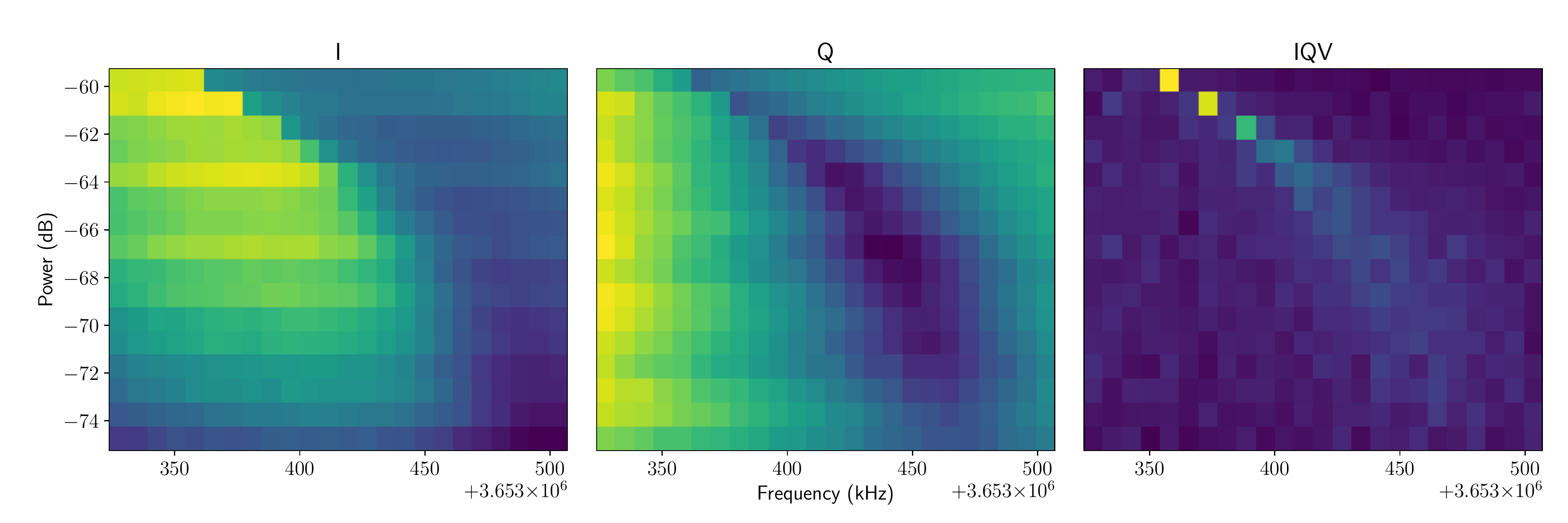}
    \caption{Example $S_{21}$ images of I, Q, and IQV around a resonator. This is before any normalization or other preprocessing, so units/scaling is arbitrary. The classification score image assigned to this input data is plotted in figure \ref{fig:wpsmap_res}.}
    \label{fig:iqv_image}
\end{figure}

\subsection{Neural Network}

\subsubsection{Preprocessing}

A typical $S_{21}$ dataset will contain the full 4-8 GHz band, sampled at 7.7 kHz, and 31 powers sampled in 1 dB increments.  For the sliding window classifier, we use a window size of 30 points in frequency ($\approx 230\ \textrm{kHz}$) by 7 points in power. In frequency space, this provides good coverage across the full linewidth of most resonators, and in power space, it allows the transition from saturation $\rightarrow$ underpowered to be fully sampled. In addition to the $S_{21}$ $I$ and $Q$ data, we include $IQV$ as well since it is a useful visual aid when manually inspecting resonators. $I$, $Q$, and $IQV$ are implemented as separate ``color channels" in the neural network input \footnote{``color channels" can be thought of as analogous to RGB channels in conventional images}.

We perform the following normalizations on each input window:
\begin{enumerate}
    \item Normalize by $S_{21}$ magnitude, independently at each power:
        \begin{itemize}
            \item $M_p = \sqrt{\frac{1}{N_{freq}}\sum_{i}{I_{p,i}^2 + Q_{p,i}^2}}$, where $p$ indexes power and $i$ indexes frequency. $N_{freq}$ is the window size along the frequency axis.
            \item $I_p \rightarrow I_p/M_p$; $Q_p \rightarrow Q_p/M_p$; $IQV_p \rightarrow IQV_p/M_p$;
        \end{itemize}
        The value of $M_p$ depends on how the MKID readout ADCs and DACs were configured for the frequency sweep at that particular power $p$; this can change arbitrarily with power and has no physical relevance. Normalizing separately at each $p$ removes this nonuniformity.
    \item Center the data, also independently at each power: $I_p \rightarrow I_p - mean_i(I_p)$, $Q_p \rightarrow Q_p - mean_i(Q_p)$, $IQV_p \rightarrow IQV_p - mean_i(IQV_p)$

\end{enumerate}

\noindent Normalizing by $M_p$ before centering ensures that the relative loop size (i.e. the amount it attenuates the probe tone on resonance) is preserved; this is an important metric of resonator quality \cite{} \footnote{loop size is proportional to $Q/Q_c$, where $Q$ is the resonator quality factor and $Q_c$ is the coupling quality factor; ideal resonators have $Q \approx Q_c$, hence a relative loop size close to 1 \cite{}}. If a window overlaps with the boundaries of the $S_{21}$ image, it is edge-padded to the correct size. 

\subsubsection{Network Architecture}

\begin{figure}[h]
    \centering
    \includegraphics[scale=.54, trim={94 50 35 70}, clip]{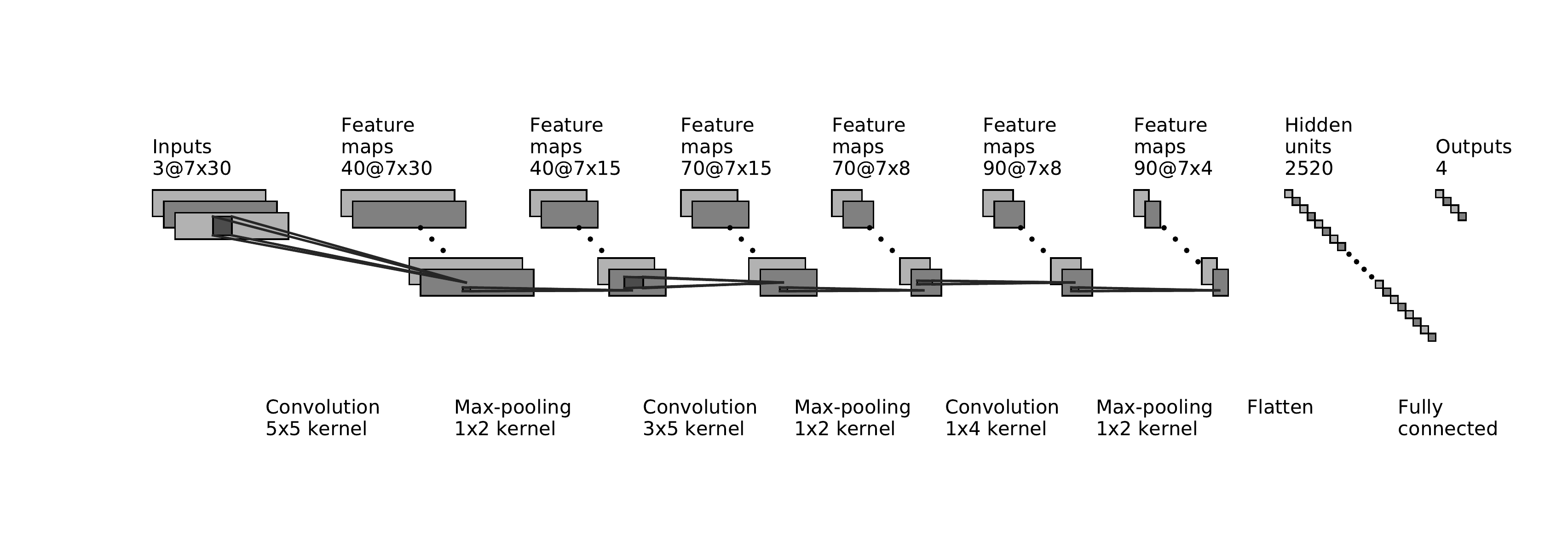}
    \caption{Schematic of neural network architecture, showing alternating convolutional/pooling layers, followed by a fully connected output layer. The row axis in each layer is along the power dimension, and the column axis is frequency. Batch normalization and a RELU activation function are applied to the output of each convolution. During training, dropout is applied to the output of each max-pooling layer. Each of the four outputs represents the classification score for each of the four possible classes. This figure was generated by adapting the code from \url{https://github.com/gwding/draw_convnet}.}
    \label{fig:nn_schematic}

\end{figure}

The convolutional neural network is implemented using the TensorFlow \cite{tensorflow} package version 1.8. The network has three convolutional/pooling layers followed by a fully connected output layer (figure \protect\ref{fig:nn_schematic}). Each convolutional/pooling layer has three stages: 1) 2D convolution (generalized across input/output color channels) with filter coefficients determined during training; 2) Rectilinear activation function, which applies $RELU(x) = max(0, x)$ to each pixel in the output of the convolution\protect\cite{imagenet}; 3) max-pooling, which downsamples the output of the previous stage by selecting the maximum value within a $n \times m$ window. The fully connected layer multiplies the flattened output of the final convolutional/pooling layer by a matrix of trainable weights to arrive at the final four element output vector. A softmax function is applied to this output to normalize the vector of classification scores. During training, batch normalization is applied to the output of each convolution to improve performance/convergence time \cite{batchnorm}, and dropout is used to minimize overfitting \protect\cite{dropout}.

\subsubsection{Training}
\label{sect:nn_training}

The network is trained using human-reviewed resonator data from the ``baseline methodology" analysis pipeline described in the previous section. Sixteen training images are made for each identified resonator (four per class):
\begin{enumerate}
        \item Four copies centered around correct power/frequency (``good" class)
        \item Four ``bifurcated" resonators, from powers $[p_c + p_{sat}, p_c + p_{sat} + 3]$ 
        \item Four ``underpowered" resonators, from powers $[p_c - p_{up}, p_c - p_{up} - 3]$
        \item Four ``no resonator" images randomly selected from regions of $S_{21}$ without resonators 
\end{enumerate}
All of the above powers are in dB; $p_c$ is the correct power, and $p_{sat}$ and $p_{up}$ represent bifurcation/underpowered thresholds above/below the correct power, respectively. These thresholds are arbitrary and can be tuned as hyperparameters. For the bifurcated training images, a random offset within $[0,  -75\ \textrm{kHz} \times \frac{f_{res}}{4\ \textrm{GHz}}]$ is added to the resonant frequency; this was found to correct for resonant frequency movement with power, and prevent the network from choosing frequencies away from the $IQV$ peak during inference.

Training a neural network is the process of finding the set of convolutional filter coefficients and fully connected layer weights that minimize a ``loss function" that is computed across all of the training data. We use a ``cross-entropy" \cite{} loss function given by: $L = -\sum_i\sum_c y_{i, c} log(p_{i,c})$, where $i$ indexes the training samples and $c$ indexes classes. $y_{i, c}$ represents the ``true" class label for sample $i$; $y_{i, c} = 1$ for the correct class $c$ and 0 for all other $c$. $p_{i, c}$ is the score assigned by the neural network for image $i$ and class $c$. Because the final layer of the neural network is a softmax function, $0 < p_{i, c} < 1$. For the training process, we use the Adam optimization algorithm, which is a modified version of stochastic gradient descent \cite{adam}, with a batch size of 50 training images and the learning rate left as a tunable hyperparameter.

\subsection{Resonator Identification}
\label{sec:nn_resid}

\begin{figure}[h]
    \centering
    \includegraphics[scale=0.39, trim={140, 0, 50, 38}, clip]{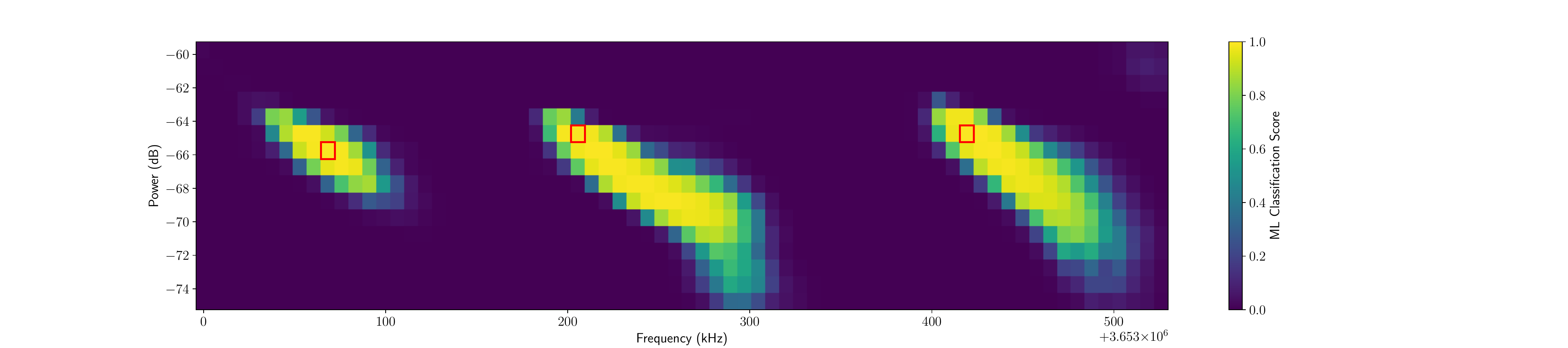}
    \caption{Output classification image for the ``good resonator'' class over a 500 kHz band. Each of the three local maxima (outlined in red) corresponds to a resonator. The input data corresponding to the rightmost resonator is plotted in figure \ref{fig:iqv_image}.}
    \label{fig:wpsmap_res}
\end{figure}

After running the full $S_{21}$ inference dataset through the neural network, we obtain a four-color $N_{power} \times N_{freq}$ output image containing the classification scores for each data point in the input image (figure \ref{fig:block_diagram}). For the purposes of inference, we only use the first ``color" of the output image, which corresponds to the likelihood of a resonator at the correct drive power and frequency being centered on a given point $(p,f)$. Correctly driven resonators correspond to local maxima in this output layer above a certain threshold (figure \ref{fig:wpsmap_res}). We find that this threshold varies between datasets, so it is generally sufficient to pick the $N_{res}$ highest local maxima, where $N_{res}$ is the expected number of resonators. In practice, we use both of these criteria to limit false positive detections (i.e. we pick the $N_{res}$ highest local maxima but require them to be above a certain threshold). For a typical 4-6 GHz sweep, we use $N_{res} = 1024$, since this is the maximum number of pixels supported by the readout system over this bandwidth \cite{readout}. It is also very close to the target pixel count for both DARKNESS \cite{darkness} ($N = 1000$) and MEC \cite{mec} ($N = 1022$).

\begin{figure}[h]
    \centering
    \includegraphics[scale=0.39, trim={140, 0, 50, 38}, clip]{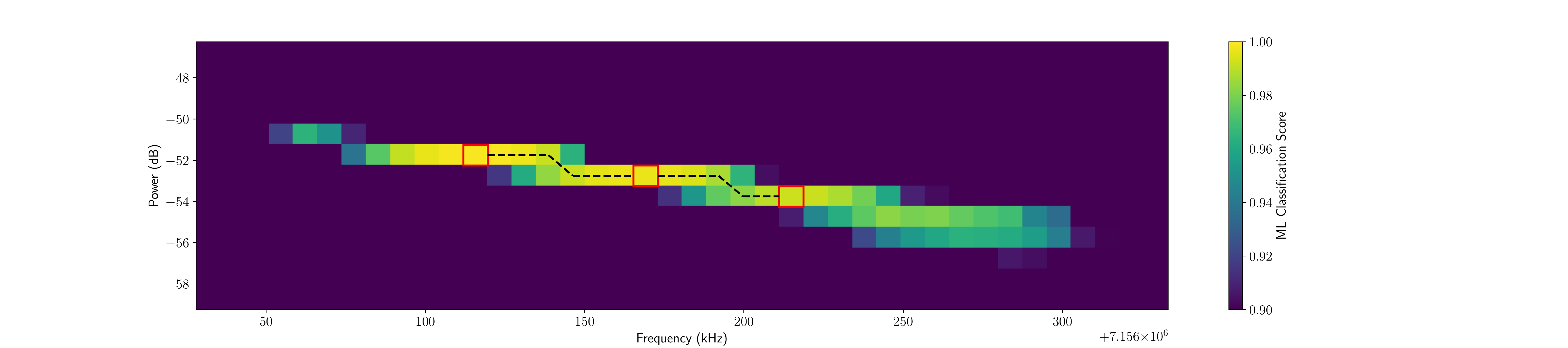}
    \caption{Output classification image for the ``good resonator'' class for a single resonator. The resonator frequency changes significantly enough with power that the resonator generates three distinct local maxima with high classification scores ($>0.99$), outlined in red. The shallowest paths between these maxima are shown by the dashed lines. The minimum value along each of these paths is very high ($>0.98$), so these maxima are not ``prominent'' enough to indicate the presence of three separate resonators. So, the algorithm will only flag the highest scoring peak as a resonator and discard the other two.}
    \label{fig:wpsmap_pr}
\end{figure}

One issue with this approach is that if a resonator's frequency changes a significant amount with power, the resonator may show up as multiple distinct local maxima. One way to mitigate this is to use a slightly modified version of the notion of topographic peak prominence \footnote{Peak prominence is a measure of how ``well-separated'' a peak is from its neighbors; two neighboring resonators are more likely to manifest as distinctly separated peaks in the output classification than a single resonator moving in frequency space.} : for all adjacent (in frequency space) local maxima meeting the initial threshold or $N_{res}$ cut, we take the minimum value along the shallowest monotonic path between the two maxima (see figure \ref{fig:wpsmap_pr}). If this value exceeds a certain threshold ($t_{pr}$; see table \ref{tbl:hyperparams}), we flag the larger of the two maxima as a resonator and discard the other one.

\section{Final Model Selection and Training}
\label{sec:eval_training}

To develop, tune, and train our model, we used a dataset consisting of five feedlines from the device ``Hypatia'' (8490 resonators), which is the device currently mounted in the MEC instrument. For each feedline, the dataset spanned 4 GHz of bandwidth in steps of 7.629 kHz, and 31 dB of power in steps of 1 dB. Resonators were identified and tuned using the baseline methodology (section \ref{sec:baseline_method}), and after the automated steps, both resonant frequency and drive power were corrected by a manual operator. All hyperparameter and model architecture tuning was performed within this dataset, with all hyperparameters being manually selected (i.e. no grid search or similar algorithms were used). Once we settled on a model architecture and hyperparameter set, we used all five feedlines to train the final model for evaluation. 

\begin{figure}[h]
    \includegraphics[]{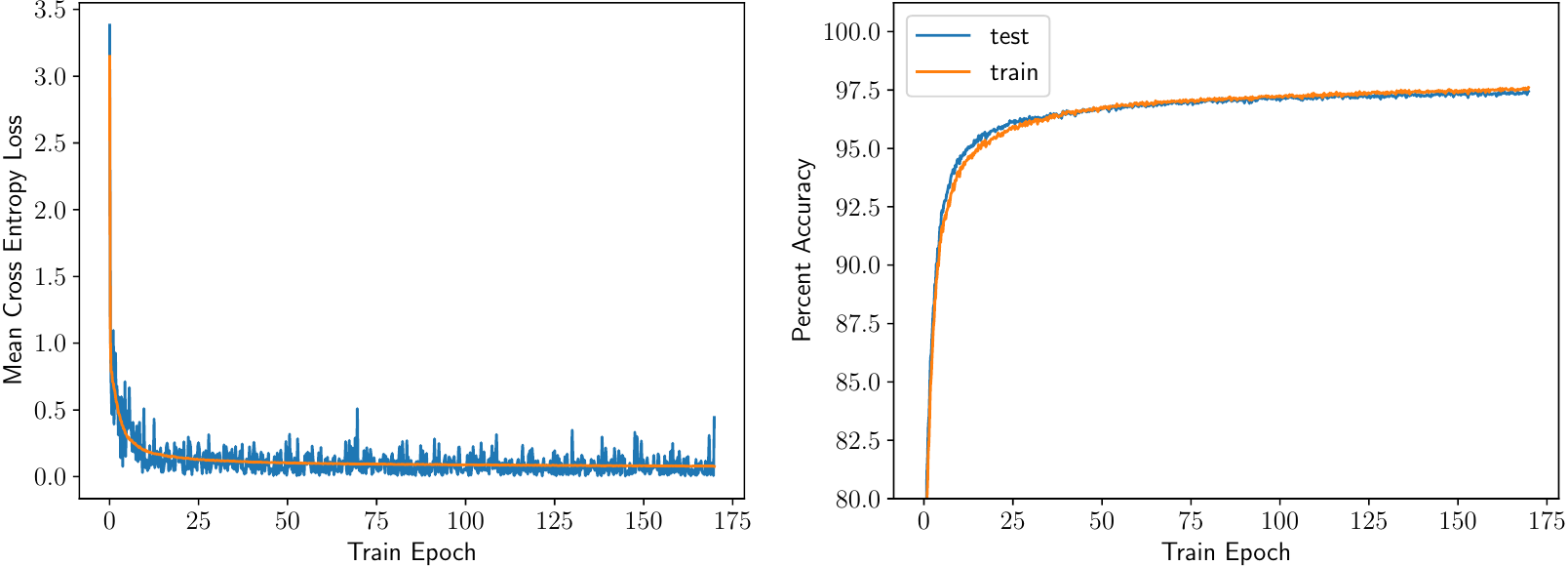}
    \caption{Mean cross entropy (left) and classification accuracy (right) for the training data and held out test images during the training process. Both of these indicators track very closely, indiating minimal to no overfitting.}
    \label{fig:train_acc_ce}
\end{figure}

During training, 10\% of the training data is held out for testing; accuracy and cross-entropy are measured throughout the training process. We do this to ensure that the neural network isn't overfitting to the training examples. After the last training step, our model has approximately 97\% accuracy on on both the training data and held-out test data (figure \ref{fig:train_acc_ce}), indicating good performance with minimal overfitting \footnote{Note that this is not a test of end-to-end system performance, as it only indicates the classification accuracy within a set of training images selected using the methods described in \ref{sect:nn_training}}.

\begin{table}[h]
    \begin{subtable}[t]{0.3\textwidth}
        \centering    
        \begin{tabular}[t]{c|c}
            Preprocessing &  \\
            \hline \hline
             $p_{sat}$ & 1 dB \\
             $p_{up}$ & 4 dB
        \end{tabular}
         
    \end{subtable}
    \hfill
    \begin{subtable}[t]{0.3\textwidth}
        \centering 
        \begin{tabular}[t]{c|c}
            Training & \\
            \hline \hline
            learning rate & $10^{-4.5}$ \\
            training epochs & 170 \\
            batch size & 50 \\ 
            dropout probability & 0.4 \\
        \end{tabular}
    \end{subtable}
    \hfill
    \begin{subtable}[t]{0.3\textwidth}
        \centering 
        \begin{tabular}[t]{c|c}
        Identification & \\
        \hline \hline
        threshold score & 0.9 \\
        $N_{res}$ & 1024 \\
        $t_{pr}$ & 0.85 \\
        \end{tabular}
    \end{subtable}
    \caption{Table of model hyperparameters used during data preprocessing, training, and identification. Neural network hyperparameters can be found in figure \ref{fig:nn_schematic}. }
    \label{tbl:hyperparams}
\end{table}

\section{Performance Evaluation}

To evaluate full system performance, we used our trained model to perform inference on a ``fresh'' $S_{21}$ test dataset, which we never used to train, develop, or tune our model. This dataset consists of a single feedline (call this Elson0) from a different MEC array than the one we used in section \ref{sec:eval_training}, which should provide a good test of model generalization. The dataset has the same span and step size as the training data in both the frequency and power dimensions. To provide a basis for comparison, we also performed resonator identification and tuning using the ``baseline methodology'', and had two different manual operators tune the drive power of each resonator. We evaluated our model on three different performance criteria, which we detail below: resonator identification, drive power tuning, and frequency placement.

\subsection{Resonator Identification}
\label{sec:eval_res_id}

For both the baseline methodology and end-to-end ML pipeline, we evaluated the number of true positive, false positive, and false negative identifications. To determine the true and false positive values, we had a manual operator screen all of the resonators identified by each method and flag all false identifications. Since we don't have a ``ground truth'' list of all of the resonators in Elson0, we estimate the number of false negatives of each method by comparing against the other method; i.e. if a resonator is found by the baseline method but not by the end-to-end ML method, it is considered a false negative of end-to-end ML. To correlate resonators across methods for this analysis, we matched each resonator from the baseline method frequency list to its unique nearest neighbor within 200 kHz in the end-to-end ML method frequency list. Unmatched resonators remaining in each list after this matching process were labeled as false negative identifications for the other method. The results of this analysis are detailed in table \ref{tbl:resid}. The end-to-end ML and baseline methods have similar performance on this metric, although the ML algorithm appears to be slightly more sensitive.

\begin{table}[h]
\begin{center}
\begin{tabular}{r|c c}
            & Baseline Method & End-to-end ML \\
            \hline
     True Positive & 1840 & 1857 \\
     False Positive & 11 & 14 \\
     False Negative & 62 & 45 \\

\end{tabular}
\end{center}
\caption{Resonator identification performance on Elson0. The dataset spans 4 Ghz of bandwidth over a single feedline, so in the absence of fabrication errors we expect at most 2044 resonators. All resonators included in this table satisfy the ``usable resonator'' criteria defined in section \ref{sec:baseline_power}.}
\label{tbl:resid}
\end{table}

\subsection{Drive Power Tuning}

\begin{figure}
    \includegraphics[]{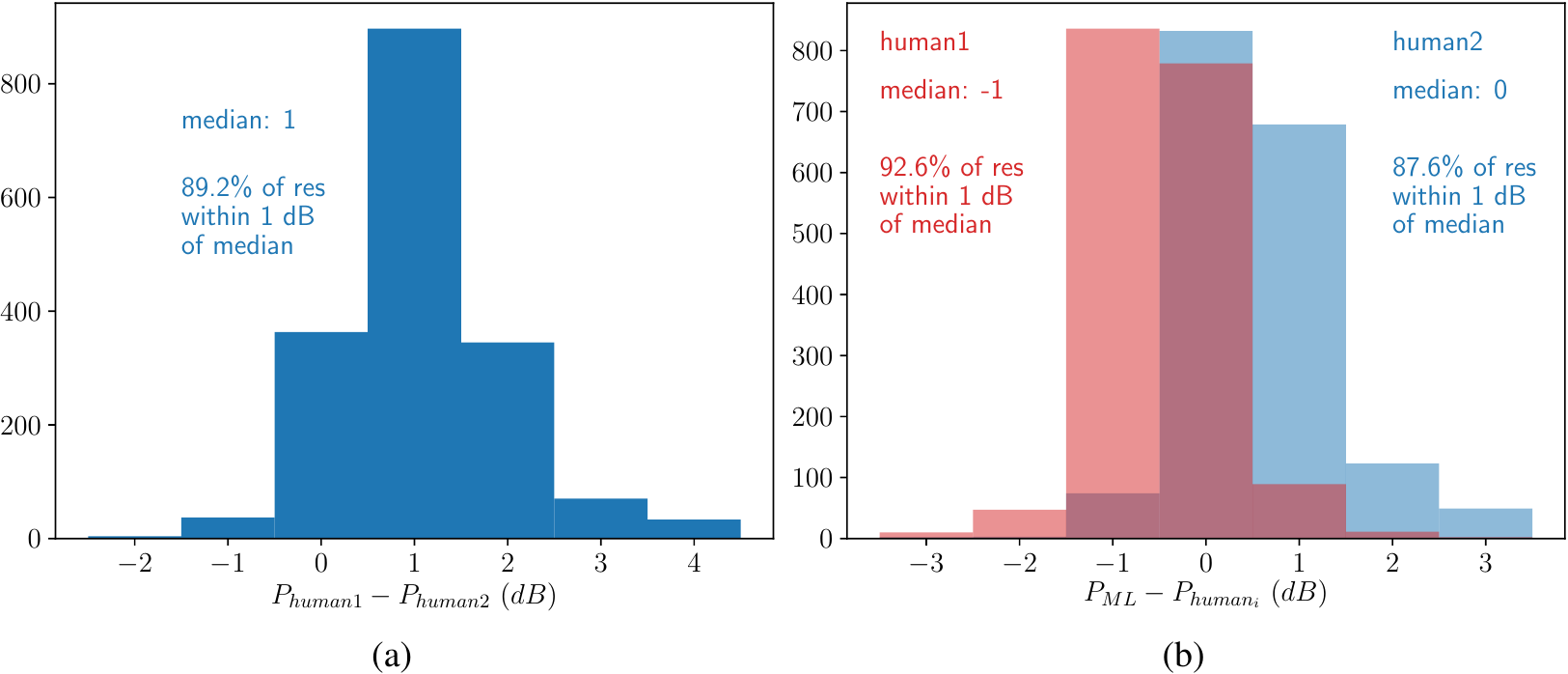}
    \caption{Histograms of the per-resonator difference in selected resonator drive power between different tuning methods for the Elson0 dataset. a) Difference between the two human operators; b) Difference between the end-to-end ML method and each human operator. For these comparisons, resonators were correlated between frequency lists using the same method as in section \ref{sec:eval_res_id}. Power differences are quantized to 1 dB (as per the dataset), which sets the histogram bin spacing}
    \label{fig:power_plots}
\end{figure}

There is some uncertainty in the manual selection of resonator drive powers; since it is done heuristically by-eye \cite{dodkins}, different operators can assign the same resonator different drive powers and each might be considered ``correct''. This uncertainty is reflected in the training data, and will fundamentally limit the performance of our pipeline. To quantify this uncertainty and provide a benchmark for model performance, we had two different human operators (call these human1 and human2) tune the drive power of each resonator in the test dataset. In figure \ref{fig:power_plots}a, we histogram the difference in selected drive power for each resonator between the two manual operators. We find that human1 on average assigns a power approximately 1 dB higher than human2, with $\sigma = 1.4$ dB. In light of this uncertainty, we expect a well performing ML algorithm to behave as a third human operator; i.e. the distributions of power differences between it and each of the human operators should be comparable to figure \ref{fig:power_plots}a. We histogram these distributions in figure \ref{fig:power_plots}b. We find these three distributions to be broadly similar; each has a systematic offset (i.e. median) $\leq 1$ dB, and a similar spread, with ${\approx}90\%$ of resonators falling within 1 dB of the median drive power difference. 

\begin{figure}
    \centering
    \includegraphics[scale=.7, trim={0 0 0 40}, clip]{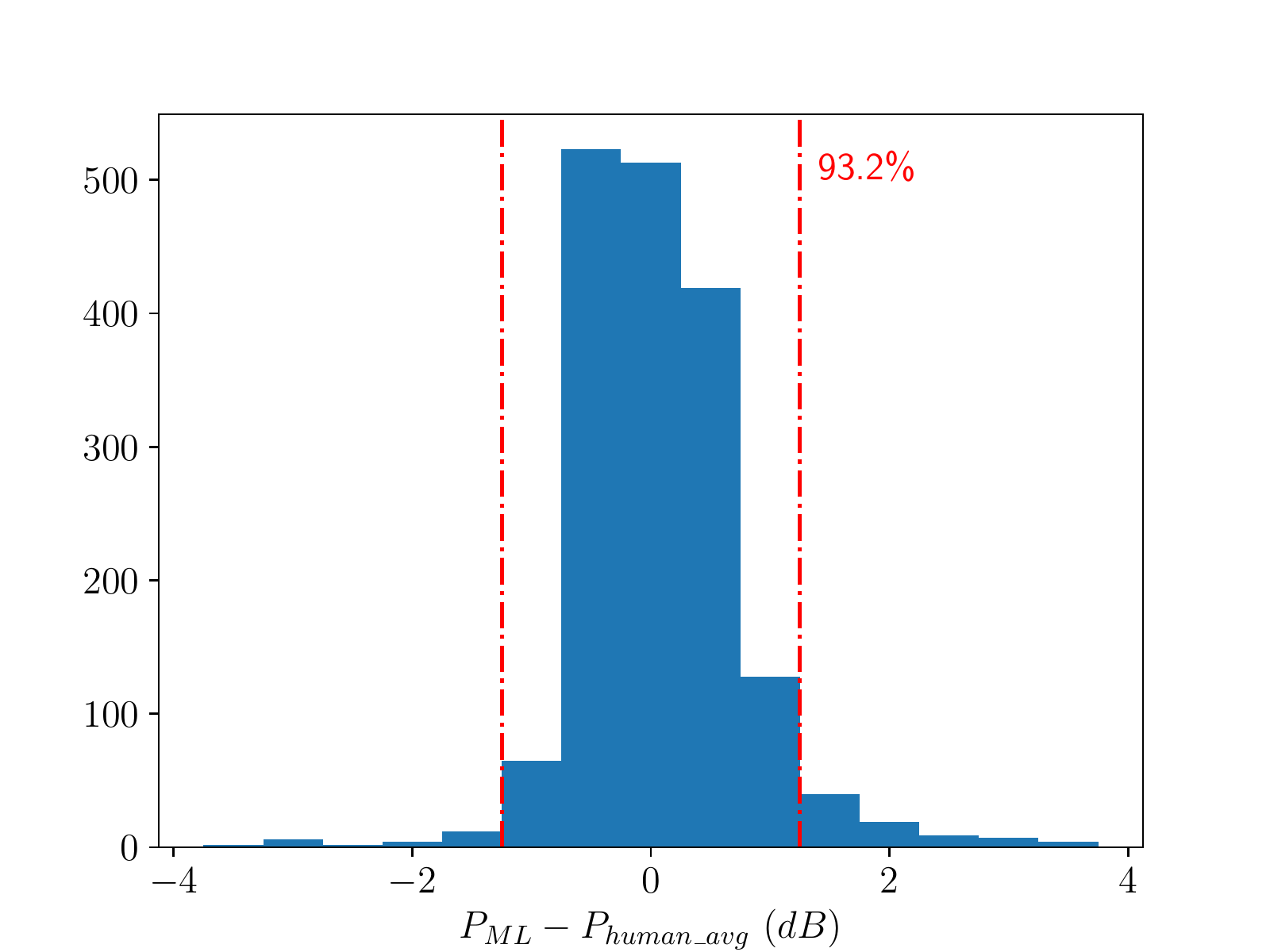}
    \caption{Histogram of the per-resonator difference in selected drive power between end-to-end ML and the average human-selected drive power for Elson0. The end-to-end ML method assigns 93\% of the resonators a drive power within 1 dB of the human average.}
    \label{fig:avg_power_plot}
\end{figure}

We can also perform an analysis similar to the one in (Ref. \citeonline{dodkins}). For this analysis, we define a ``ground truth'' power for each resonator by averaging together the powers assigned to it by each human operator. We then define a per-resonator binary accuracy metric: a resonator was assigned the correct drive power by the ML algorithm if and only if $|P_{ML} - P_t| \leq 1$, where $P_{ML}$ and $P_t$ are the ML assigned and ``ground truth'' powers in dB, respectively. By this metric, our pipeline accurately classifies 93\% of the Elson0 resonators (figure \ref{fig:avg_power_plot}). This is comparable to the 90\% accuracy achieved by the ML algorithm (and manual operators!) in (Ref. \citeonline{dodkins}); both are likely limited by training data nonuniformity.

\subsection{Frequency Placement}
\label{sec:freq_place}

\begin{figure}[h]
    \centering
    \includegraphics[scale=.7, trim={0 0 0 40}, clip]{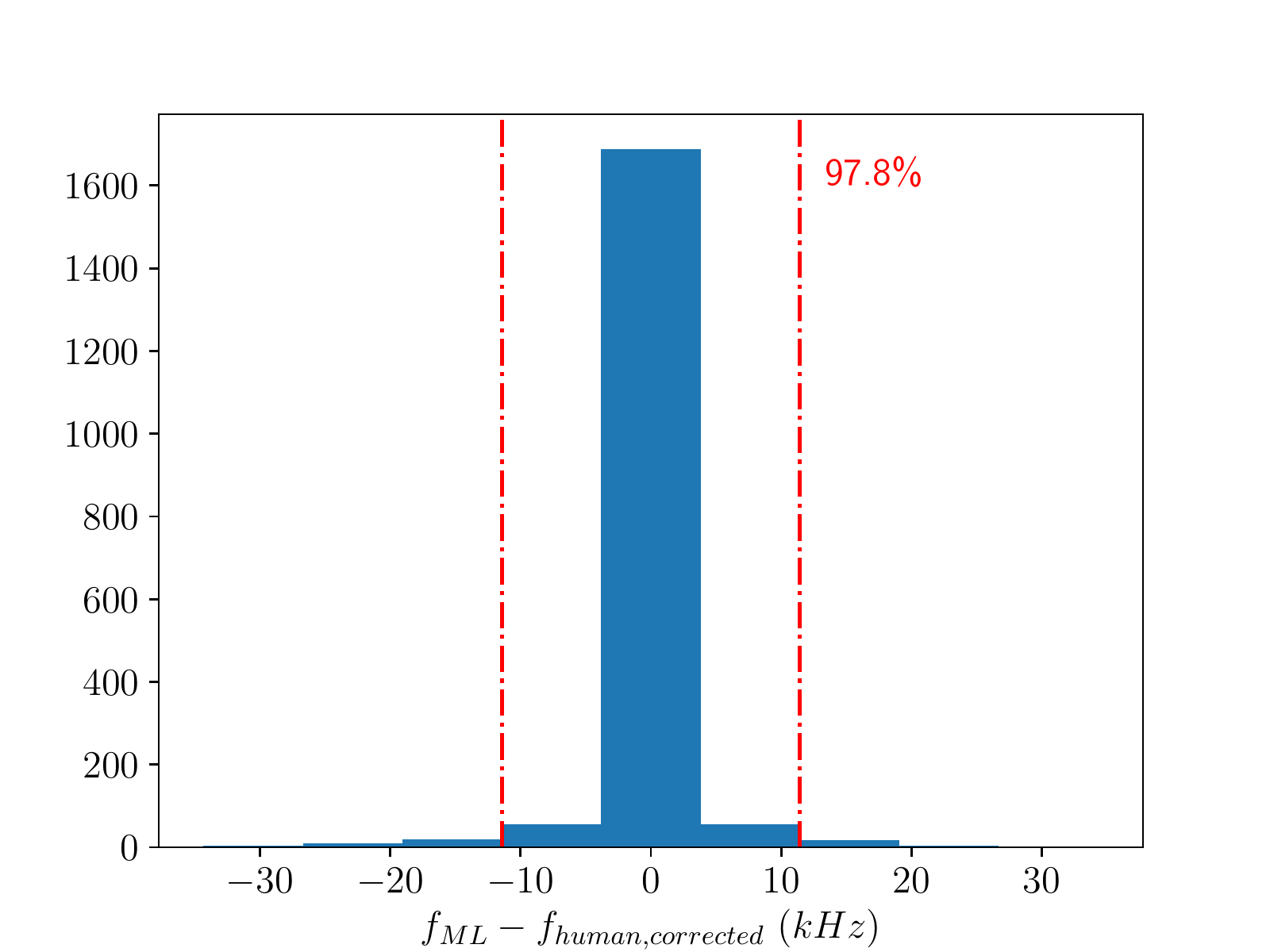}
    \caption{Histogram of difference between ML selected frequency and any human correction (performed by a single human operator). Frequency differences are quantized to 7.629 kHz, which sets the histogram bin spacing. 90.4\% of resonators required no correction, and 97.8\% of resonators were within one quantization step of the corrected frequency.}
    \label{fig:freq_perf_plot}
\end{figure}

Since the resonator frequency shifts significantly with power, the ideal frequency placement depends on resonator drive power. So, we cannot simply compare the ML selected frequency with the baseline solutions. Instead, we had a human operator check through the end-to-end ML solutions and correct any misplaced resonator frequencies. These results are plotted in figure \ref{fig:freq_perf_plot}. 97.8\% of resonators were assigned a frequency within 7.629 kHz of the correct value; this represents an error of less than 10\% of the typical resonator linewidth, so the impact on performance is likely negligible. 

\subsection{Computational Performance}

Using three desktop CPU cores (Intel Core i5), along with an NVIDIA RTX-2080 GPU, our pipeline takes approximately 12 minutes to calibrate a full feedline. Image generation and normalization forms the bulk of this time. 

\section{Conclusion}

We have developed a deep learning based analysis pipeline, which, for the first time allows us to completely automate the calibration of optical/IR MKID array frequency combs. Our pipeline is in active use, reducing the 40+ hour manual tuning process for the MEC instrument array to less than two hours of computational time. This has signficantly streamlined the array characterization process, and reduced instrument downtime in the event of a warm-up to room temperature (which often causes resonators to shift in frequency). It is also a crucial step towards feasibly scaling up MKID array pixel counts. 

When calibrated using our method, MEC has a demonstrated energy resolution $R = 3.8$ and 75\% pixel yield on working feedlines ($R = 4.8$ and the yield is 82\% on the well-filled ``sweet spot'' of the array)\cite{mec}. See (Ref. \citeonline{mec}) for more details; all array performance metrics provided were obtained using an array calibrated by our pipeline.

The precision of the calibration solutions (particularly resonator drive power) is inherently limited by the lack of consistency of the human-generated training data. Future work will aim to address this; potential solutions involve using analytical fitting methods to generate or refine training data.

Our pipeline is open source and can be found here: \url{https://github.com/MazinLab/MKIDReadout}.

\section{Acknowledgements}

We would like to thank Nicholas Zobrist, Isabel Lipartito, Sarah Steiger, Noah Swimmer, Jenny Smith, Kristina Davis, and Clint Bockstiegel for spending many tedious hours to help generate the training and validation data used in this work. We would also like to thank Bryan He for valuable discussions about CNN architecture selection/validation. Neelay Fruitwala is supported by a grant from the Heising-Simons Foundation.


\bibliography{report}   

\begin{thebibliography}{10}

\bibitem{paul2017}
P.~Szypryt, S.~R. Meeker, G.~Coiffard, {\em et~al.} {\em Opt. Express} {\bf
  25}, 25894  (2017).

\bibitem{mec}
A.~B. Walter, N.~Fruitwala, S.~Steiger, {\em et~al.}, ``The {MKID} exoplanet
  camera for {Subaru} {SCExAO},'' {\em Publications of the Astronomical Society
  of the Pacific} {\bf 132}(1018), 125005  (2020).

\bibitem{darkness}
S.~R. Meeker, B.~A. Mazin, A.~B. Walter, {\em et~al.}, ``{DARKNESS}: a
  microwave kinetic inductance detector integral field spectrograph for
  high-contrast astronomy,'' {\em Publications of the Astronomical Society of
  the Pacific} {\bf 130}(988), 065001  (2018).

\bibitem{swenson}
L.~Swenson, P.~Day, B.~Eom, {\em et~al.}, ``Operation of a titanium nitride
  superconducting microresonator detector in the nonlinear regime,'' {\em
  Journal of Applied Physics} {\bf 113}(10), 104501  (2013).

\bibitem{zobrist}
N.~Zobrist, B.~H. Eom, P.~Day, {\em et~al.}, ``Wide-band parametric amplifier
  readout and resolution of optical microwave kinetic inductance detectors,''
  {\em Applied Physics Letters} {\bf 115}(4), 042601  (2019).

\bibitem{dodkins}
R.~Dodkins, S.~Mahashabde, K.~O’Brien, {\em et~al.}, ``{MKID} digital readout
  tuning with deep learning,'' {\em Astronomy and Computing} {\bf 23}, 60--71
  (2018).

\bibitem{scipy}
P.~Virtanen, R.~Gommers, T.~E. Oliphant, {\em et~al.}, ``{{SciPy} 1.0:
  Fundamental Algorithms for Scientific Computing in Python},'' {\em Nature
  Methods} {\bf 17}, 261--272  (2020).

\bibitem{readout}
N.~Fruitwala, P.~Strader, G.~Cancelo, {\em et~al.}, ``Second generation readout
  for large format photon counting microwave kinetic inductance detectors,''
  {\em Review of Scientific Instruments} {\bf 91}(12), 124705  (2020).

\bibitem{deeplearning}
Y.~LeCun, Y.~Bengio, and G.~Hinton, ``Deep learning,'' {\em nature} {\bf
  521}(7553), 436--444  (2015).

\bibitem{tensorflow}
M.~Abadi, P.~Barham, J.~Chen, {\em et~al.}, ``Tensorflow: A system for
  large-scale machine learning,'' in {\em 12th {USENIX} Symposium on Operating
  Systems Design and Implementation ({OSDI} 16)},  265--283, {USENIX}
  Association, (Savannah, GA)  (2016).

\bibitem{imagenet}
A.~Krizhevsky, I.~Sutskever, and G.~E. Hinton, ``Imagenet classification with
  deep convolutional neural networks,'' {\em Communications of the ACM} {\bf
  60}(6), 84--90  (2017).

\bibitem{batchnorm}
S.~Ioffe and C.~Szegedy, ``Batch normalization: Accelerating deep network
  training by reducing internal covariate shift,'' {\em arXiv preprint
  arXiv:1502.03167}   (2015).

\bibitem{dropout}
N.~Srivastava, G.~Hinton, A.~Krizhevsky, {\em et~al.}, ``Dropout: a simple way
  to prevent neural networks from overfitting,'' {\em The journal of machine
  learning research} {\bf 15}(1), 1929--1958  (2014).

\bibitem{adam}
D.~P. Kingma and J.~Ba, ``Adam: {A} method for stochastic optimization,'' in
  {\em 3rd International Conference on Learning Representations, {ICLR} 2015,
  San Diego, CA, USA, May 7-9, 2015, Conference Track Proceedings},  Y.~Bengio
  and Y.~LeCun, Eds.  (2015).

\end{thebibliography}
\bibliographystyle{spiejour}   


\vspace{2ex}\noindent\textbf{Neelay Fruitwala} attended the California Institute of Technology, and graduated in 2015. He is currently a graudate student at the University of California at Santa Barbara, focusing on the development and application of Microwave Kinetic Inductance Detector (MKID) instrumentation for exoplanet direct imaging.

\vspace{2ex}\noindent\textbf{Prof. Benjamin A Mazin} attended Yale University, graduating in 1997.  He then attended the California Institute of Technology, graduating with a doctorate in Astrophysics in August, 2004.  After a short post-doc at Caltech, he went to work as a scientist at JPL in March, 2005. He joined the faculty at the University of California Santa Barbara in September, 2008, where he leads a lab dedicated to the development of optical/UV/X-ray Microwave Kinetic Inductance Detectors (MKIDs) and astronomical instrumentation for time and energy resolved studies. 

\vspace{1ex}
\noindent Biographies and photographs of the other authors are not available.

\listoffigures
\listoftables

\end{spacing}
\end{document}